\documentclass[conference,usenatbib]{basi}
\usepackage[T1]{fontenc}
\usepackage[british]{babel}
\usepackage[varg]{txfonts}
\usepackage{rotating}
\usepackage{dcolumn}
\begin{document}
\title[Author running title]{Spectral and Temporal Properties of MAXI J1836-194 during 2011 Outburst}
\author[Jana et~al.]%
       {A.~Jana$^1$\thanks{email: \texttt{argha@csp.res.in}},
       D.~Debnath$^{1}$, S. Mondal$^1$, S.~K.~Chakrabarti$^{1,2}$, A. A. Molla$^1$, D. Chatterjee$^1$\\
       $^1$ Indian Centre for Space Physics, 43 Chalantika, Garia Station Road, Kolkata, 700084, India\\
       $^2$ S. N. Bose National Centre for Basic Sciences, JD-Block, Salt Lake, Kolkata, 700098, India}
\pubyear{2015}
\volume{12}
\pagerange{137-138}
%\status{submitted}

\date{Received --- ; accepted ---}

\maketitle
%------------------------------------------------------------------------------%
% abstract and keywords                                                        %
%------------------------------------------------------------------------------%
\label{firstpage}

\begin{abstract}
We study black hole candidate (BHC) MAXI~J1836-194 during its 2011 outburst with 
Two Component Advective Flow (TCAF) model using RXTE/PCU2 data in $2.5-25$~keV band. 
From spectral fit, accretion flow parameters such as Keplerian disk rate ($\dot{m_d}$), sub-Keplerian halo rate 
($\dot{m_h}$), shock location ($X_{s}$) and compression ratio (R) are extracted directly. 
During the entire phase of the outburst, quasi-periodic oscillations (QPOs) are observed sporadically. 
From the nature of the variation of accretion rate ratio (ARR=$\dot{m_h}$ / $\dot{m_d}$) and QPOs, 
entire period of the outburst is classified in two spectral states, 
such as, hard (HS), hard-intermediate (HIMS).
Unlike other transient BHCs, no signature of soft (SS) and soft-intermediate (SIMS) spectral states are observed 
during entire phase of the outburst.

\end{abstract}


%------------------------------------------------------------------------------%
% main text of the paper, using \section, \subsection, \subsubsection          %
%------------------------------------------------------------------------------%
\section{Introduction}\label{s:intro}

The transient BHC MAXI~J1836-194 was discovered simultaneously by MAXI/GSC and SWIFT/BAT on 
2011 Aug. 29 at sky location of R.A. $= 18^h35^m43.43^s$, Dec $= -19^\circ 19'12.1''$.
This is a short orbital period ($< 4.9~hrs.$) and highly rotating BHC (spin parameter $a=0.88\pm0.03$) 
with mass and distance are predicted in the range of $4-12~M_\odot$, $4-10$~kpc respectively 
(see, Jana et al., 2016 for references). 
%Cenko et al. (2011) predicted the companion as a high massive Be star. 
Recently after the inclusion of TCAF model (Chakrabarti \& Titarchuk, 1995) into NASA's spectral analysis software package 
XSPEC as an additive table model, accretion flow dynamics of few BHCs (GX~339-4, H~1743-322, MAXI~J1659-152) 
are well understood (see, Debnath et al., 2014, 2015a,b; Mondal et al., 2014). 
This motivated us to study current outburst of MAXI~J1836-194 with the model (see, Jana et al., 2016).

%------------------------------------------------------------------------------%
% example figures to illustrate various styles...
%------------------------------------------------------------------------------%

%\section{Data Analysis}\label{s:fonts}

\begin{figure}
\vskip -0.3cm
\centerline{\includegraphics[scale=0.6,width=6.5truecm,angle=0]{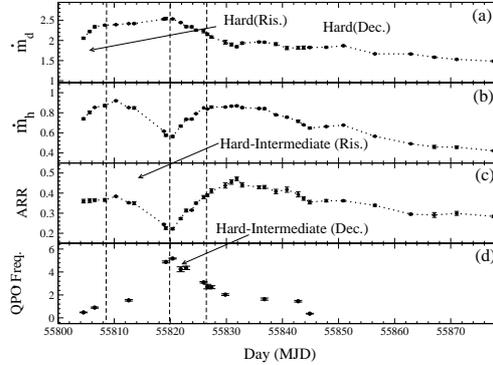}}
\vskip -0.5cm
\caption{Variations of TCAF fitted/derived (a) Keplerian disk ($\dot{m_d}$) rate, (b) sub-Keplerian halo ($\dot{m_h}$) rate,
(c) ARR ($\dot{m_h}$/$\dot{m_d}$), and (d) observed QPO frequency with day (MJD) are shown.}
\end{figure}

\section{Result and Discussion}

We analyze RXTE/PCA data for 35 observations, spread over entire 2011 outburst. $2.5-25$~keV PCU2 spectra are fitted 
with TCAF in XSPEC %(by freezing mass of the BH at $9.5~M_\odot$) 
to extract physical flow parameters, such as, 
$\dot{m_d}$ in Eddington rate, $\dot{m_h}$ in Eddington rate, $X_s$ in Schwarzschild radius and $R$. 
%QPOs are observed sporadically on and off during the entire phase of the outburst. 
Based on the variation of ARR and nature of QPOs, entire phase of the 2011 outburst of MAXI~J1836-194 is 
classified in the following sequence: HS (ris) $\rightarrow$ HIMS (ris) $\rightarrow$ HIMS (dec) $\rightarrow$ HS (dec). 
During the first five observations, the BHC was in hard state with clear dominance in $\dot{m_h}$. 
On 2011 Sept 6 (MJD=55810.29), ARR attains maximum as we observe a transition from HS to HIMS (ris).
On Sept 16 (MJD=55820.41), the BHC enters in HIMS (dec), when highest QPO frequency of 5.17~Hz is observed.
%After that it enters in HIMS (declining).
%Then a transition from HIMS to SIMS is observed on 2011 Sept 14 (MJD=55818.85), 
%when ARR reaches its minimum value (with maxima in $\dot{m_{d}}$). Then after $\sim 2$ days, it enters into the HIMS of declining phase. 
On 2011 Sept 27 (MJD=55831.85) ARR reaches its maximum value (with the maxima in $\dot{m_h}$) and the source enters into the HS (dec). 
Strangely SS and SIMS are missing during the entire phase of the outburst. The reason behind this may be, the BH is immersed 
into the excretion disk of a high massive companion Be star. % and surrounded by a the low angular momentum wind matter from the companion. 
%So, most of the time during this outburst, the source was flooded with low-angular momentum optically thin sub-Keplerian component. 
%Also no monotonic evolution of QPO frequencies during rising and declining HS and HIMS are missing. 
We do not find QPOs on regular basis, although trends of monotonically increasing (0.47-5.17~Hz) and decreasing (5.17-0.37~Hz) 
nature of QPO frequencies are observed during rising and declining phases respectively of the outburst (Fig. 1d).
%\textbf{In the rising phase, we find QPO in 3 observations.
%In the HIMS (dec) and HS (dec) phase, we observe QPO sporadically with the frequency in monotonically decreasing nature.
This is may be due to non-satisfaction of resonance condition between cooling and infall time scales (Chakrabarti et al., 2015; Mondal et al., 2015).

%------------------------------------------------------------------------------%
%\section*{Acknowledgements}
%
%I would like to thank Prof. S.K. Chakrabarti, Dr. Dipak Debnath, Mr. Santanu Mondal
%Mr. A.A. Molla for their help and support in this project. I would also like to thank
%ISRO Respond project. I also thank RETCO-II for giving me opportunity to present my work.

%------------------------------------------------------------------------------%

\begin{thebibliography}{}

\bibitem[Chakrabarti \& Titarchuk(1995)Chakrabarti \& Titarchuk]{CT95} Chakrabarti, S.K., \& Titarchuk, L.G., 1995, ApJ, 455, 623
\bibitem[Chakrabarti et al.(2015)Chakrabarti et al.]{C15} Chakrabarti, S.K., \& Mondal, S., \& Debnath, D., 2015, MNRAS, 452, 3451
%\bibitem[Debnath et al.(2008)Debnath et al.]{DD08} Debnath, D., Chakrabarti, S.K., \& Nandi, A., et al., 2008, BASI, 36, 151
%\bibitem[Debnath et al.(2010)Debnath et al.]{DD10} Debnath, D., Chakrabarti, S. K., \& Nandi, A., 2010 A$\&$A, 520, 98
%\bibitem[Debnath et al.(2013)Debnath et al.]{DD13} Debnath, D., Chakrabarti, S.K., \& Nandi, A., 2013, AdSpR, 52, 2143
\bibitem[Debnath et al.(2014)Debnath et al.]{DCM14} Debnath, D., Mondal, S., \& Chakrabarti, S.K., 2014, MNRAS, 440, L121
\bibitem[Debnath et al.(2015a)Debnath et al.]{DMC15} Debnath, D., Mondal, S., \& Chakrabarti, S.K., 2015a, MNRAS, 447, 1984
\bibitem[Debnath et al.(2015b)Debnath et al.]{DMCM15} Debnath, D., Molla, A.A., Chakrabarti, S.K., \& Mondal, S., 2015b, ApJ, 803, 59
\bibitem[Jana et al.(2016)Jana et al.]{Jana16} Jana, A., Debnath, D., \& Chakrabarti, S.K., et al., 2016, ApJ, 819, 107
\bibitem[Mondal et al.(2014)Mondal et al.]{MDC14} Mondal, S., Debnath, D., \& Chakrabarti, S.K., 2014, ApJ, 786, 4
\bibitem[Mondal et al.(2015)Mondal et al.]{MCD15} Mondal, S., Chakrabarti, S.K., \& Debnath, D., 2015, ApJ, 798, 57

\end{thebibliography}
\end{document}